\documentclass[reprint,amsmath,aps,prb,footinbib]{revtex4-2}
\usepackage{graphicx,epstopdf,dcolumn,bm,mathtools,siunitx,multirow,hyperref}
\begin{document}
\title{Quantum shape oscillations in the thermodynamic properties \\ of confined electrons in core-shell nanostructures}
\thanks{NOTICE: This is an author-created, un-copyedited version of an article accepted for publication in Journal of Physics: Condensed Matter. IOP Publishing Ltd is not responsible for any errors or omissions in this version of the manuscript or any version derived from it. The Version of Record is available online at https://doi.org/10.1088/1361-648X/ac303a.}
\author{Alhun Aydin}
\email{alhunaydin@fas.harvard.edu}
\affiliation{Department of Chemistry and Chemical Biology, Harvard University, Cambridge, MA 02138, USA}
\author{Jonas Fransson}
\author{Altug Sisman}
\affiliation{Department of Physics and Astronomy, Uppsala University, 75120, Uppsala, Sweden}
\date{\today}
\begin{abstract}
Quantum shape effect appears under the size-invariant shape transformations of strongly confined structures. Such a transformation distinctively influences the thermodynamic properties of confined particles. Due to their characteristic geometry, core-shell nanostructures are good candidates for quantum shape effects to be observed. Here we investigate the thermodynamic properties of non-interacting degenerate electrons confined in core-shell nanowires consisting of an insulating core and a GaAs semiconducting shell. We derive the expressions of shape-dependent thermodynamic quantities and show the existence of a new type of quantum oscillations due to shape dependence, in chemical potential, internal energy, entropy and specific heat of confined electrons. We provide physical understanding of our results by invoking the quantum boundary layer concept and evaluating the distributions of quantized energy levels on Fermi function and in state space. Besides the density, temperature and size, the shape per se also becomes a control parameter on the Fermi energy of confined electrons, which provides a new mechanism for fine tuning the Fermi level and changing the polarity of semiconductors. 
\end{abstract}
\maketitle
\section{Introduction}
Physical properties of nanostructures are affected by their geometry when the thermal de Broglie wavelength of particles is comparable with the sizes of the structure. As a manifestation of this geometry dependence, quantum size effects constitute one of the backbones of nanoscience and nanotechnology, paving the way to the enhancement of electrical, optical, thermal and thermoelectric properties of materials  \cite{baltes,mitchen,rodun,tebook2014}.

Quantization of energy levels in low-dimensional nanosystems causes oscillations in the electronic properties of materials. The well-known examples are the quantum oscillations in resistivity (Shubnikov–de Haas effect) and in magnetization (de Haas–Van Alphen effect) caused by Landau quantization due to the magnetic field \cite{Shoenberg_1984}. Energy level quantization due to size effects also causes quantum oscillations in certain thermodynamic and transport properties \cite{McIntyre_1976,Champel_2001,Rogacheva_2009,Grenier_2013,Rogacheva_2015,qsetransporteir,Rogacheva_2017,aydin6,Rogacheva_2019,Tsysar_2020} of various semiconductor and metallic nanostructures which have been extensively studied due to their importance on nanoscale electronics and nano-engineered devices  \cite{Trivedi_1988,Volokitin_1996,Yoffe_2002,Weis_2017}. In low dimensional materials, quantum size effects \cite{rodun,Dai_2004,sismanmuller} become stronger and give rise to distinct subbands \cite{bineker} and size quantization effect \cite{Kenkre_1972,Nenadovic_1985}. Stepwise (staircase-like) behaviors of some physical quantities (\textit{e.g.} internal energy \cite{aydin1}, conductance \cite{bineker}) and oscillatory characteristics of others (\textit{e.g.} specific heat \cite{Holstein_1973,DEOLIVEIRA2004424,TAYURSKII2003152,Boyacioglu_2012,aydin1}, thermopower \cite{Rogacheva_2005,Fust_2019}) appear due to the nature of Fermi-Dirac distribution function and its derivative with respect to energy (so called occupancy variance or thermal broadening function) respectively  \cite{aydinhvm,aydin4,aydin5}. Size-dependent quantum oscillations attracted a great deal of interest particularly in recent decades \cite{McIntyre_1976,Rogacheva_2009,_zer_2009,Grenier_2013,Rogacheva_2015,qsetransporteir,Rogacheva_2017,aydin6,Rogacheva_2019,Tsysar_2020}.

Sizes of a nanostructure are determined by geometric size parameters (Weyl parameters) \cite{baltes,aydinphd} \textit{i.e.} volume $V$, surface area $A$, peripheral lengths $P$ and number of vertices $N_V$ under the standard Lebesgue measure. Here $A,P$ and $N_V$ constitute the lower-dimensional sizes which can play a significant role in lower-dimensional systems. Changing the size parameters of a domain also causes change of its shape and vice versa in general. However, recently it has been shown that distinguishing the size and shape effects from each other is possible through a size-invariant shape transformation, which enables changing the shape of a nanostructure without altering its geometric size parameters. The process of size-invariant shape transformation elicits a new physical phenomenon called the \textit{quantum shape effect} \cite{aydin7,aydinphd,aydin11} at nanoscale.

In this paper, we show the existence of distinct, shape-dependent quantum oscillations in the thermodynamic quantities of electrons due to the quantum shape effect (QShE). We investigate the chemical potential, internal energy, entropy and specific heat of electrons in core-shell nanowires. We examine the shape-dependent quantum oscillations for various temperature and electron densities (concentrations), ranging from weakly to moderately degenerate regimes. In particular, the oscillatory behavior of chemical potential and internal energy induced by the QShE is interesting since those quantities do not oscillate but exhibit essentially stepwise behavior in case of the variations in size \cite{aydin1}. The role of shell volume in core-shell nanostructures has not been well-understood before \cite{Sirigu_2017}. Within the perspective of the quantum boundary layer (QBL) concept \cite{qbl,uqbl,aydin7,aydinphd}, we provide physical insights on the role of shell volume in the shape-dependent thermodynamic properties. In particular, the shell volume becomes an effective volume due to the quantum nature of confined electrons and plays an important role in the existence of the new type of quantum oscillations. We investigate the distributions of quantized energy levels on Fermi function and in state space to explain the origins of these oscillations. Quantum shape dependence of chemical potential opens the possibility of tuning the Fermi energy or degeneracy by controlling the shape parameter. We also find that while the oscillation amplitudes are higher for lower electron densities and temperatures in general, there are deviations from this general trend. The deviations are due to the fact that density dictates the magnitude of the QShE more dominantly than temperature. Furthermore, the frequency of quantum shape oscillations increases with increasing electron density while their amplitudes become weaker. We see that the magnitude of oscillations due to QShE is in the order of that of quantum size effects for the chosen system parameters here. Finally, we show the oscillatory violation of entropy-heat capacity equivalence at the high degeneracy limit due to distinct shape dependencies of both quantities.

In the following section, Sec. \ref{II}, we provide our model, formalism and semi-analytical expressions. We begin Sec. \ref{III} by presenting and examining the quantum shape oscillations. We then discuss the masking of the temperature effect, oscillatory violation of entropy-heat capacity equivalence and mention the possibility of controlling the Fermi energy by varying the shape. We conclude our findings in Sec. \ref{IV}.

\section{Model and Formalism}\label{II}
\subsection{Non-interacting electrons in a core-shell nanostructure}
We consider a core-shell nanowire structure with an insulating core and semiconducting shell. The schematic transverse views of the considered core-shell nanowires is seen in Fig. 1. QShE takes place in the shell nanowire, when the core nanowire is rotated along its transverse axis. Notice that such a transformation preserves all of the geometric size variables, while still allowing the change of the shape of shell structure where electrons are confined. Naturally, rather than actual rotation of the core structure, core-shell nanowires could also be thought to be prepared separately so that properties of nanowires with the different angular configurations can be compared with each other.

\begin{figure}[b]
\centering
\includegraphics[width=0.48\textwidth]{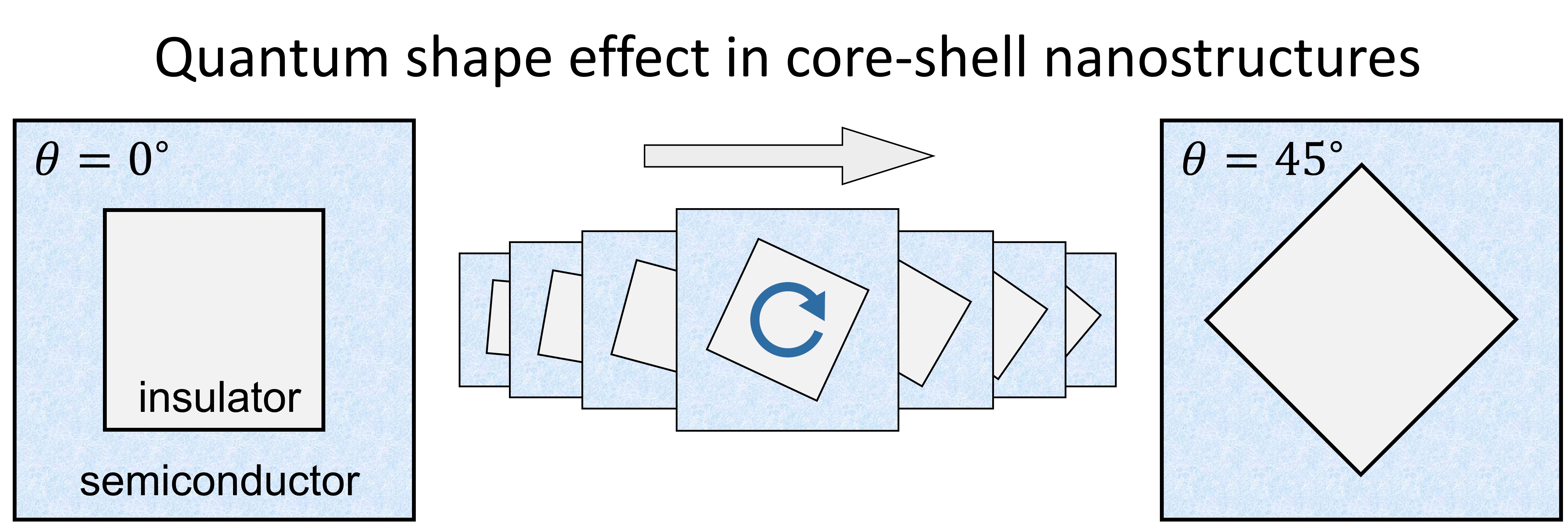}
\caption{Transverse views of core-shell nanowires. Rotation of the core wire leads to a size-invariant shape transformation of the shell structure wherein the electrons are confined. Quantum shape effect is associated with and controlled by the independent variable $\theta$.}
\label{fig:pic1}
\end{figure}

We are interested in the thermodynamic properties of electrons confined within the shell structure, where the change of domain shape occurs and is controllable by the variable $\theta$, denoting the rotation angle in the transverse axis. In this regard, semiconductor core-shell nanostructures are suitable materials for QShE to occur \cite{aydin11,Musin_2006,Yang_2008,Tsuji_2008,An_2013,Dimitriadis_2015,Manolescu_2016,Rajadell_2017,Urbaneja_Torres_2018,Jahan_2019}. Thus, to highlight the QShE, we choose a well-known semiconducting material of electron-doped Gallium Arsenide (GaAs) as a shell structure, due to its low effective electron mass, $m_{\text{eff}}=0.067m_e$ where $m_e$ is the bare electron mass \cite{bineker,Sze_2006}. GaAs is a direct band gap semiconductor with a spherical Fermi surface and perfectly parabolic band structure at Gamma point which justifies the use of free electron model with the effective mass approximation for the equilibrium properties of conduction electrons around band bottom\cite{Ashcroft1976,Grundmann_2010}. In particular, the non-interacting electron model is widely used for studying the equilibrium properties of GaAs nanostructures \cite{Mork_tter_2015,Arab_2016,Degtyarev_2017,Kovalenko_2020}. This model, albeit it is simple, can capture the essential geometric effects in the energy spectrum which is what we need for QShE. Note that the characteristic behaviors obtained in this study will remain almost the same but only their magnitudes will change even if we use different effective mass values. The core structure can be made of any insulating material. Side lengths of shell and core nanowires are chosen as $L_s=64$nm and $L_c=41$nm respectively whereas the longitudinal length of core-shell nanowires is $L_l=763$nm. The confinement domain with these size parameters imposes strong quantum confinement conditions in the transverse direction and very weak confinement in longitudinal direction for the electrons of GaAs in conduction band. Nanowires are chosen to be free of any defect and roughness. The presence of defects in shell structure or roughness on the boundaries could impact the results if the size of the imperfectness is in the order of Fermi wavelength (see Table 1) or even larger. In the case of small defects or roughness, however, QShE remains the dominant effect since they are induced by the entire boundaries. The effect of any imperfections can also be distinguished and separated from QShE by invoking their different temperature dependencies.

The main target of the study is to propose and examine the shape-induced quantum oscillations in thermodynamic properties of confined electrons in core-shell semiconductor structures. A reliable but simple model based on ideal conditions provides an effective tool to predict the ultimate values of these oscillations and reveal their fundamental mechanisms. Such a model gives some primary results guiding further studies besides the intuitions to see if these oscillations give us some additional and useful control possibilities on material properties. One of the most widely used approaches for calculating material properties is the density functional theory (DFT). On the other hand, it requires limitations for the number of atoms to be considered due to heavy computational loads. Here the considered domain contains tens of thousands of atoms for each atomic layer in the transverse direction. Therefore, the problem is far beyond the practical capabilities of a DFT calculation. However, the effect we consider here is mainly based on the shape-induced density variations near the Fermi surface. Therefore, albeit a full DFT simulation may change some aspects of the energy bands deeply below the Fermi level, they do not influence the physics close to the Fermi level. This is similar to the concepts of the Fermi liquid theory, for instance, where merely the properties of the electrons near the Fermi level are regarded as relevant. Hence, it is justified to restrict our discussion to the electronic density in the conduction band which is sufficiently dilute for the pertinent exchange and correlation potentials in the DFT simulations to be negligible. As a consequence, the theory is effectively single electron-like and we can just as well employ an effective mass single electron (non-interacting) Schr\"{o}dinger equation for reaching the same qualitative results. There may naturally be some small differences in magnitudes when comparing the numbers, however, on the qualitative scale, there should be no considerable differences between the methods.

We use the usual statistical mechanical framework to calculate the thermodynamic quantities of non-interacting electrons inside the shell wire. Fermi-Dirac distribution function reads $f=1/[\exp(\tilde{\varepsilon}-\Lambda)+1]$ where  $\Lambda=\mu/(k_BT)$ with $\mu$ is chemical potential and $\tilde{\varepsilon}=\varepsilon/(k_BT)$ with $\varepsilon$ is energy eigenvalues of the confined electrons with $k_B$ Boltzmann constant and $T$ temperature. Essentially, the shape dependence is embedded into the energy eigenvalues, $\{\varepsilon_1(\theta), \varepsilon_2(\theta), \varepsilon_3(\theta), \ldots\}$. Due to the strong confinement in the transverse direction, the wave nature of electrons becomes dominant and causes an increase in the discreteness of energy eigenvalues. These eigenvalues of electrons in the shell structure are obtained by numerically solving the Schr\"{o}dinger equation separately for each integer degree of angular configuration between $0^{\circ}\leq\theta\leq 45^{\circ}$. When necessary, we use even higher resolution $\Delta\theta=0.25^{\circ}$ for angular steps, especially for highly degenerate cases. To maximize the shape effect, we impose Dirichlet boundary conditions so that electrons cannot leak through the material boundaries or into the insulating part. We use COMSOL\textsuperscript{\textregistered} Multiphysics software to solve the time-independent Schr\"{o}dinger equation and obtain the energy eigenvalues for the electrons confined in the regions denoted by blue color in Fig. 1.

The elongated geometry of the nanowire allows us to make use of a bounded continuum approximation for the component of energy eigenvalues in the longitudinal direction which reduces the numerical workload from 3D to 2D workspace. Electrons are relatively unconfined (almost free) in longitudinal direction so that the number of Fermi wavelengths fitting into the length of the nanowire is large, $\lambda_F/L_l<<1$. Therefore, continuum-based analytical methods can be used to calculate the contributions coming from the longitudinal modes. Energy eigenvalues of this system can be separated as $\varepsilon=\varepsilon_t(\theta)+\varepsilon_l(L_l)$ where $\varepsilon_t$ and $\varepsilon_l$ denotes transverse and longitudinal eigenvalues respectively. By this way, we calculate transverse eigenvalues numerically, which becomes much easier for a 2D domain and obtain longitudinal contribution using the bounded continuum approach. This provides not only a quite accurate approximation, but also makes it possible to obtain semi-analytical expressions for the thermodynamic properties of electrons confined at nanoscale.

For weakly confined systems, bounded continuum approximation gives more accurate results than the usual continuum approximation, because the former takes the non-zero value of the ground state into account, whereas the latter considers a continuous spectrum starting from zero energy values \cite{baltes,pathbook,aydin2,aydin3,aydin8}. Although both approximations are sufficient for the chosen length of longitudinal direction in this work, we choose to use bounded continuum approximation to get more precise results.

\begin{figure*}[t]
\centering
\includegraphics[width=0.98\textwidth]{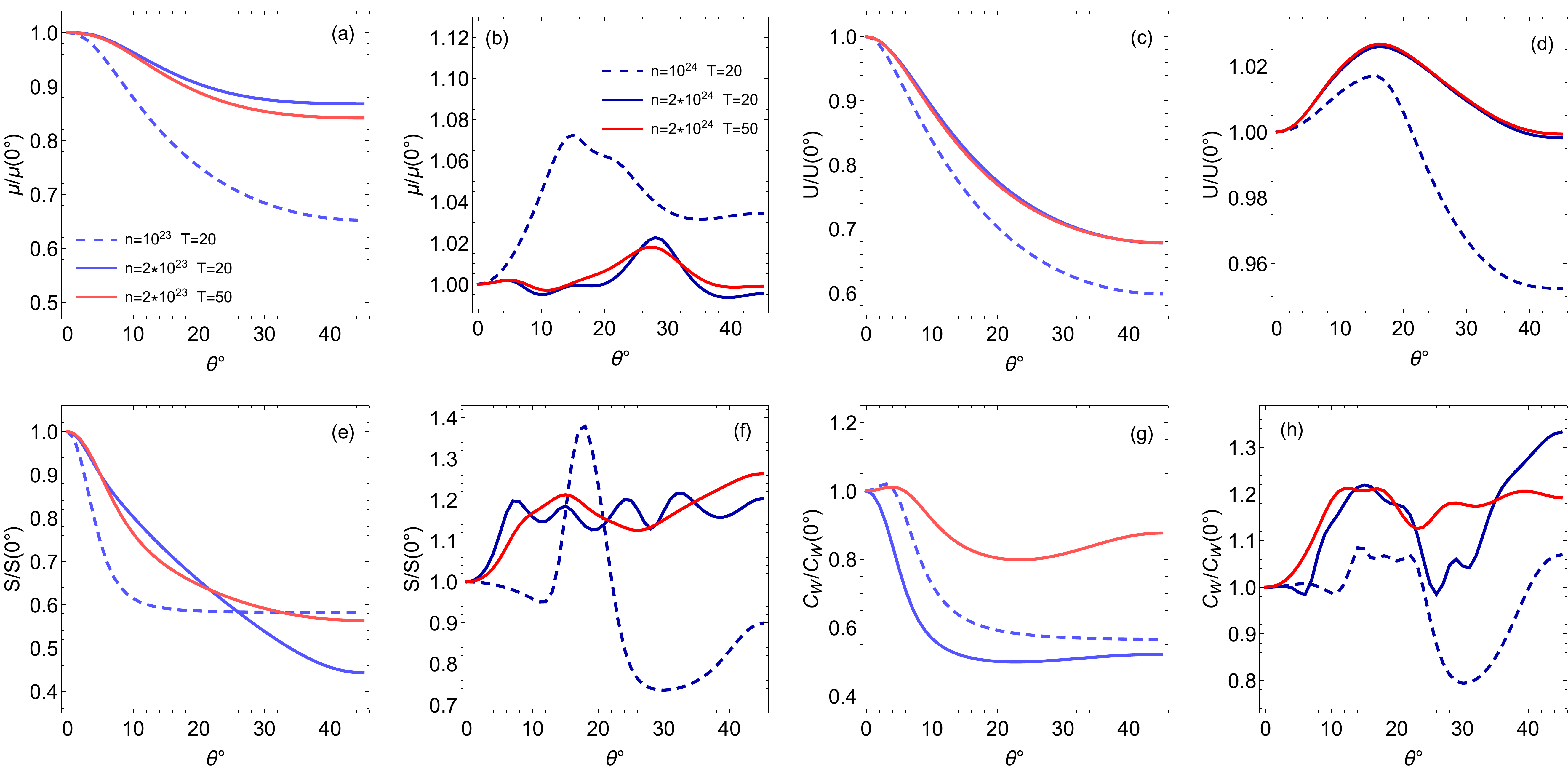}
\caption{Shape dependencies (characterized by $\theta$) of the normalized (a), (b) chemical potential, (c), (d) internal energy, (e), (f) entropy and (g), (h) heat capacity at constant Weyl parameters. Red and blue colors correspond to $T=50$K and $T=20$K temperature values respectively. Solid curves correspond to twice the electron density of those with dashed curves in their respective subfigures. (a), (c), (e) and (g) plots are for weakly degenerate, (b), (d), (f) and (h) plots are for moderately degenerate conditions, which are determined by the electron densities. All quantities are normalized to their values at $\theta=0^{\circ}$. Temperatures are given in Kelvin scale and the unit of the densities is m$^{-3}$ in the legend.}
\label{fig:pic2}
\end{figure*}

\subsection{Thermodynamic expressions}
The number of particles is determined by the summation of the distribution function over all eigenvalues and considering the spin degree of freedom $g_s$ (which is a factor of two), $N=g_s\sum_{\varepsilon} f_{\varepsilon}$. We can write $N=g_s\sum_{\varepsilon_t}\sum_{\varepsilon_l} f_{\varepsilon}$ to explicitly show the contributions of transverse and longitudinal parts. Now by invoking the first two terms of the Poisson summation formula \cite{baltes,sismanmuller} we apply the bounded continuum approximation to the summation for the longitudinal part. The number of particles gives $N\approx g_s[\sum_{\varepsilon_t}\int{fdi_l}-f(0)/2]$ where $i_l$ is the momentum state variable for longitudinal direction, noting that $\varepsilon_l=h^2i_l^2/8m_{\text{eff}}L_l^2$. Hence, the expression for the number of particles in our system is obtained as
\begin{equation}
\begin{split}
N=g_s\sum_{\varepsilon_t}{[-\frac{L_l}{\lambda_{th}}Li_{\frac{1}{2}}(z)}+\frac{1}{2}Li_{0}(z)],
\end{split}
\end{equation}
where $Li$ is the polylogarithm function with argument $z=-\exp(\Lambda-\tilde{\varepsilon}_t)$. $\lambda_{th}=h/\sqrt{2\pi m_{\text{eff}}k_BT}$ is the thermal de Broglie wavelength of unbounded electrons with effective mass $m_{\text{eff}}=0.067m_e$. The chemical potential can be numerically solved from Eq. (1) as the total number of electrons in the shell structure is equal to the density multiplied by the volume.

By the same approach, we find the semi-analytical expressions for internal energy, entropy and heat capacity respectively as follows
\begin{equation}
\begin{split}
\frac{U}{k_BT}=g_s\sum_{\varepsilon_t}[-\frac{L_l}{2\lambda_{th}}Li_{\frac{3}{2}}(z)-\tilde{\varepsilon}_t\frac{L_l}{\lambda_{th}}Li_{\frac{1}{2}}(z)+\tilde{\varepsilon}_t\frac{1}{2}Li_{0}(z)],
\end{split}
\end{equation}
\begin{equation}
\begin{split}
\frac{S}{k_B}&=g_s\sum_{\varepsilon_t}[-\frac{3L_l}{2\lambda_{th}}Li_{\frac{3}{2}}(z)+\frac{1}{2}Li_1(z)\\
&+(\Lambda-\tilde{\varepsilon}_t)\frac{L_l}{\lambda_{th}}Li_{\frac{1}{2}}(z)-(\Lambda-\tilde{\varepsilon}_t)\frac{1}{2}Li_0(z)],
\end{split}
\end{equation}
\begin{equation}
\begin{split}
\frac{C_W}{k_B}&=g_s\sum_{\varepsilon_t}{[-\frac{3L_l}{4\lambda_{th}}Li_{\frac{3}{2}}(z)}-2\tilde{\varepsilon}_t\frac{L_l}{\lambda_{th}}Li_{\frac{1}{2}}(z) \\
&-\tilde{\varepsilon}_t^2\frac{L_l}{\lambda_{th}}Li_{-\frac{1}{2}}(z)+\frac{1}{2}Li_{-1}(z) \\
&-\frac{\left[\sum_{\varepsilon_t}{-\frac{L_l}{2\lambda_{th}}Li_{\frac{1}{2}}(z)}-\tilde{\varepsilon}_t\frac{L_l}{\lambda_{th}}Li_{-\frac{1}{2}}(z)+\frac{1}{2}Li_{-1}(z) \right]^2}{\sum_{\varepsilon_t}{-\frac{L_l}{\lambda_{th}}Li_{-\frac{1}{2}}(z)}+\frac{1}{2}Li_{-1}(z)}].
\end{split}
\end{equation}
It is worthwhile to mention that here heat capacity is defined not just at constant volume but at constant Weyl parameters, which includes lower dimensional sizes as well in addition to volume. These thermodynamic expressions are valid as long as the confinement in longitudinal direction is not strong (a valid assumption for most nanowire geometries), which is also the case here.

\section{Results and Discussion}\label{III}
\subsection{Quantum shape oscillations}
We present the variations of the thermodynamic properties of electrons due to the changes in shape parameter characterized by the rotation angle of the core structure, $\theta$, in Fig. 2. All thermodynamic quantities in Fig. 2 are normalized by their respective values at $\theta=0^{\circ}$. We focus on low temperature regimes to consider the quantum degeneracy effects as well. Hot-cold color scheme by the shades of red and blue are valid for all figures where red and blue shades correspond to $T=50$K and $T=20$K respectively. Solid curves correspond to density values which are twice of the electron densities of those with dashed curves for their respective subfigures. Electron densities of $n=10^{23}$ m$^{-3}$ and $n=2\times 10^{23}$ m$^{-3}$ correspond to weakly degenerate conditions (red and blue colors in Fig. 2a, c, e, g), whereas $n=10^{24}$ m$^{-3}$ and $n=2\times 10^{24}$ m$^{-3}$ (red and blue colors in Fig. 2b, d, f, h) correspond to moderately degenerate ones. Electron density can be modified within the chosen ranges by moderately or highly doping the GaAs \cite{Sacks_1985,Arab_2016}.

Firstly, from Fig. 2, the following general inferences for the behavior of thermodynamic quantities under QShE can be drawn: (1) The magnitude of QShE (deviations due to shape variation, $\theta$) is larger in weakly degenerate regimes compared to the moderate degeneracies. (2) Quantum shape  oscillations start to appear at moderately degenerate conditions, whereas smoother changes occur for weak degeneracies. (3) The expected behavior of the usual reciprocal relation between temperature and QShE strength does not always hold.

\begin{figure*}[t]
\centering
\includegraphics[width=0.95\textwidth]{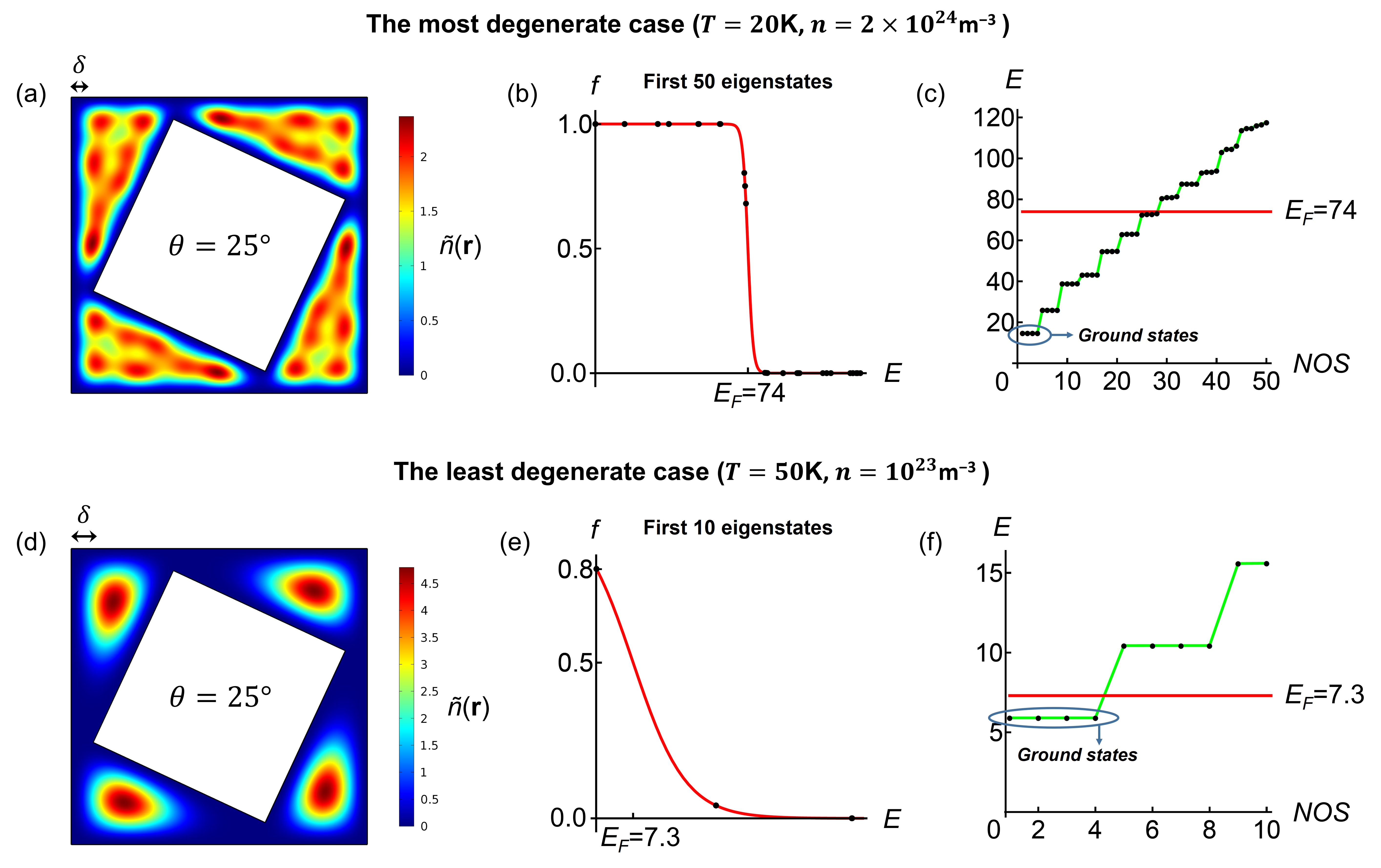}
\caption{Comparison of the quantum thermal densities (normalized to classical densities $n$) and the distribution of the states for the most degenerate case (upper row figures) and the least degenerate case (lower row figures) for the $\theta=25^{\circ}$ configuration as an example. Energies are given in the units of $k_BT$. For $T=20$K and $n=2\times 10^{24}$ m$^{-3}$, (a) Cross-sectional view of the core-shell nanowire. Appearance of Friedel oscillations in the electron density. Emergence of the effective volume and quantum boundary layers (QBLs) are also seen. $\delta$ denotes the QBL thickness. (b) Distribution of first 50 transverse energy eigenstates in the Fermi function and manifestation of a sharp Fermi surface. (c) Energy versus number of states (NOS) plot for the first 50 transverse energy eigenstates. Low-lying degeneracies of energy levels and their positions can clearly be seen. Ground state and low-lying states are usually four-fold degenerate. For $T=50$K and $n=10^{23}$ m$^{-3}$, (d) Friedel oscillations disappear in the electron density. (e) Distribution of first 10 transverse energy eigenstates in the Fermi function. Fermi surface loses its meaning (smoothening) and thermal behavior of electrons can be approximated by Boltzmann statistics. (f) Energy versus number of states plot for the first 10 energy eigenstates in the transverse direction. Degenerate ground states of transverse modes are enough to determine the shape-dependent thermodynamic behaviors in the least degenerate case.}
\label{fig:pic3}
\end{figure*}

The reason why QShE is stronger for weaker degeneracies can be understood from the electron (Fermi) wavelength analysis. The strength of the quantum shape dependence is directly related with the Fermi wavelength. Under quantum confinement, the density distribution of confined electrons becomes non-uniform near to the impenetrable domain boundaries where quantum boundary layers (QBLs) are formed in which the local density changes drastically \cite{qbl,aydin6}. Due to QBL, particles occupy an effective volume that is smaller than the apparent volume. When the confinement becomes strong, the QBLs of inner and outer cores start to overlap with each other \cite{aydin7,aydinphd}. The amount of overlap determines the strength of QShE and it is directly related with the thermal de Broglie wavelength of particles, or Fermi wavelength in this case. Therefore, in general, the larger the Fermi wavelength, the larger the influence of QShE on the system. In our previous study about QShE on thermodynamic properties of a non-interacting Maxwell-Boltzmann gas \cite{aydin7}, we were able to obtain fully analytical expressions by using the QBL method \cite{aydin7,qbl,uqbl,aydinphd}. For a brief review of the QBL method, please see Chapter 2.3 of Ref \cite{aydinphd}. In the case of degenerate Fermi gas, however, QBL becomes much more complicated due to the existence of Friedel oscillations \cite{aydin6,Jia_2010} and make it impractical to get the analytical results based on QBL approach. That's why, here we perform numerical calculations. Although the QBL concept cannot provide analytical expressions here, it helps us to interpret the underlying physical mechanisms of the quantum shape oscillations.

The temperature corrected Fermi wavelength is given by $\lambda_F(T)=2\lambda_{th}/\sqrt{\Lambda(T)}$ for unbounded particles so that $\lambda_F(T)\propto 1/\sqrt{\mu(T)}$. The comparison of Fermi wavelengths for the considered electron densities and temperatures are given in Table I. While the chemical potential is directly proportional to the Fermi energy, $\mu(T=0)$, finite temperature causes just a perturbative correction on the chemical potential under degenerate conditions, which is the case here. Finite temperature effect on electrons' wavelength is weaker than that of electron density, as is clearly seen from Table I. Due to this fact, changes in electron density have a stronger effect on the strength of QShE, compared to the changes in temperature. Sometimes this can cause the masking of the finite temperature effects on the QShE, on which we will elaborate more in the next section. In any case, it is clear that weakly degenerate conditions are more favorable for QShE because of the larger Fermi wavelengths which make the quantum confinement effects stronger.

\begin{table}[t]
\caption{Finite-temperature electron (Fermi) wavelength values for the considered electron densities and temperatures.}
\def\arraystretch{1.2}
\setlength{\tabcolsep}{0.5em}
\begin{tabular}{cccc}
\hline
Conditions                                                      & n (m$^{-3}$)      & T (K) & $\lambda_{F}$ (nm) \\ \hline
                                                                & $10^{23}$         & 20    & 22.60              \\
\begin{tabular}[c]{@{}c@{}}Weakly\\ degenerate\end{tabular}     & $2\times 10^{23}$ & 20    & 21.41              \\
                                                                & $2\times 10^{23}$ & 50    & 21.56              \\ \hline
                                                                & $10^{24}$         & 20    & 16.32              \\
\begin{tabular}[c]{@{}c@{}}Moderately\\ degenerate\end{tabular} & $2\times 10^{24}$ & 20    & 13.35              \\
                                                                & $2\times 10^{24}$ & 50    & 13.37              \\ \hline
\end{tabular}
\end{table}

For moderately degenerate conditions, on the other hand, QShE lead to an irregular oscillatory behavior, despite their weaker magnitude. Quantum oscillations due to quantum size effects at nanoscale have been attributed to the strong variations in the occupations of the states near to the Fermi surface \cite{Rogacheva_2012,aydin4,aydin5,aydinhvm,Rogacheva_2017}. A similar effect also plays a role here in the systems under QShE. However, quantum shape-dependent oscillations cannot be explained solely by the states near the Fermi surface. The oscillatory behaviors appear not only in entropy and heat capacity, but also in the chemical potential and internal energy of electrons, which do not occur under quantum size effects. Entropy and heat capacity are the quantities related with the occupancy variance  (derivative of the Fermi distribution function). Therefore, quantum oscillations in those quantities result from the variations of the states near the Fermi surface. The chemical potential and internal energy, however, are directly related with the occupancy function itself (the Fermi distribution function), not its variance. Fermi surface states have only negligible contributions on chemical potential and internal energy, yet we observe noticeable oscillatory-like behaviors in those quantities. This suggests that quantum shape oscillations seen in Fig. 2(b) and (d) have a different origin than the ones in Fig. 2(f) and (h). 

\begin{figure}[b]
\centering
\includegraphics[width=0.48\textwidth]{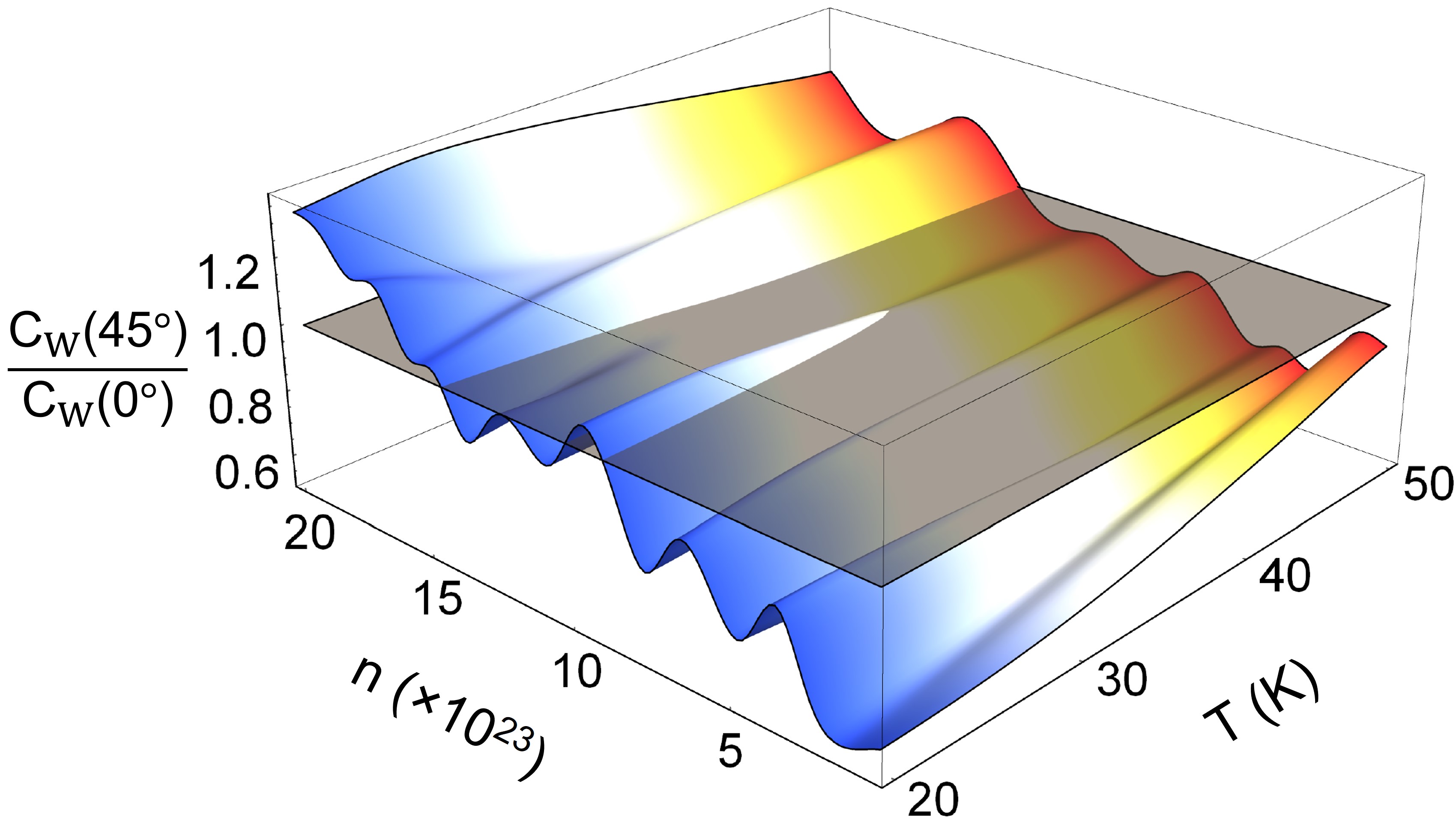}
\caption{The ratio of specific heats for $\theta=0^{\circ}$ and $\theta=45^{\circ}$ configurations varying with temperature and electron density. Grey plane denotes the zero change as a reference. Oscillatory behavior exists in every region while the amplitude of deviation grows with decreasing density and temperature in general.}
\label{fig:pic4}
\end{figure}

The origin of the internal energy oscillations lies in the chemical potential and its shape dependence. To find the chemical potential, we fix the number of particles in the system. The chemical potential does not oscillate under size variations because the occupancies of energy levels do not fluctuate but shift accordingly with the size (e.g. due to the opening of subbands). Even though the geometric (apparent) volume of the confinement domain remains unchanged under QShE, shape transformations cause an effective volume change in the confinement domain of electrons due to overlap of QBL of core and shell structures. Since we keep the number of electrons constant at each case, the change in effective volume causes a variation in the effective density and the occupancies of energy states of electrons. However, unlike the size effects, the occupancies of the states do not just shift but fluctuate \cite{aydinphd}, because of the irregular shape of the effective domain as a result of oscillatory behavior of QBL in a degenerate and confined Fermi gas \cite{aydin6}. This complicated behavior of the occupancies causes the oscillatory changes in the chemical potential as well as the quantities depending on it. Hence, there are two types of oscillations, one is the usual quantum oscillations due to the fluctuations of the states near Fermi surface and a second type of oscillation related to the fluctuations of Fermi level itself as a result of fluctuating effective volume induced by QShE. 

\begin{figure*}[t]
\centering
\includegraphics[width=0.9\textwidth]{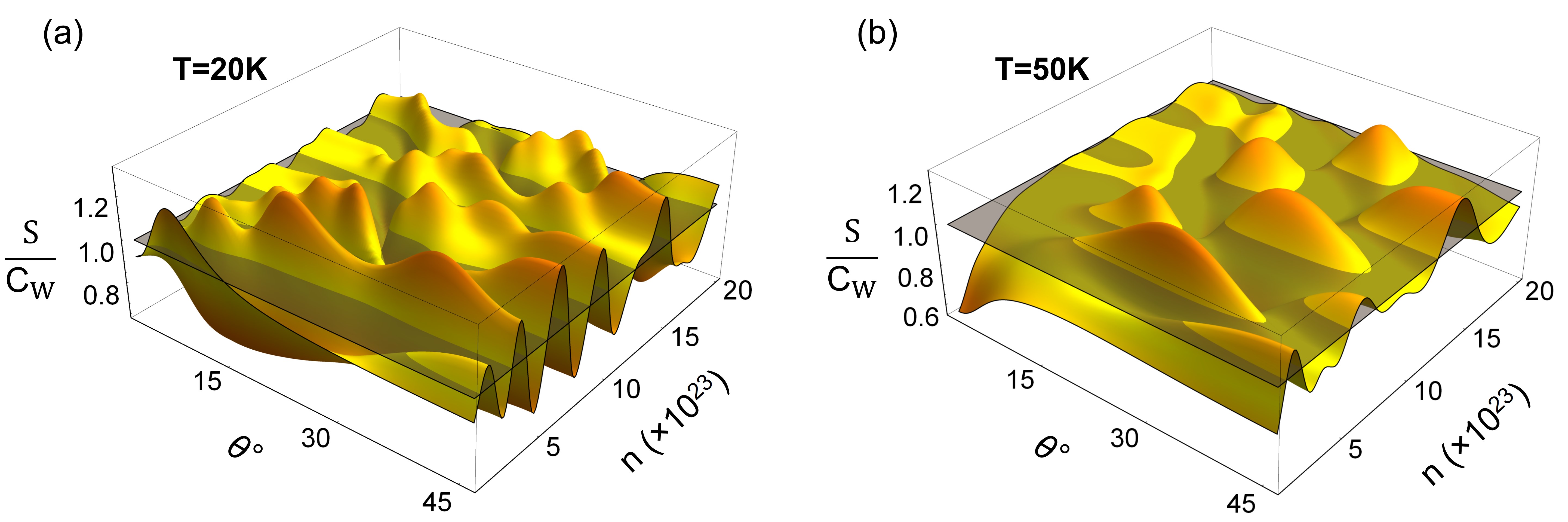}
\caption{Oscillatory violation of entropy-heat capacity equivalence due to the changes in shape and electron density at (a) 20K and (b) 50K temperatures. Violation is stronger at the peaks denoted by darker shades of orange. Grey plane represents the entropy-heat capacity equivalence as a reference.}
\label{fig:pic5}
\end{figure*}

To make the discussions more clear and explain the underlying physics in detail, we compare the two most distinct cases that are considered here: the most degenerate ($T=20$K and $n=2\times 10^{24}$ m$^{-3}$) and the least degenerate ($T=50$K and $n=10^{23}$ m$^{-3}$) cases. We pick the $\theta=25^{\circ}$ configuration as an example. In Fig. 3(a) we plot the cross-sectional view of the local density of electrons in the shell structure at thermal equilibrium (quantum thermal density). The normalized quantum thermal density is given by
\begin{equation}
\tilde{n}_q(\mathbf{r})=\frac{n_q(\mathbf{r})}{n}=\frac{\sum_{\varepsilon}f_{\varepsilon}\left|\Psi_\varepsilon(\mathbf{r})\right|^2}{\frac{1}{V}\sum_{\varepsilon}f_{\varepsilon}},
\end{equation}
where $n_q$ denotes the ensemble averaged local probability density, $n$ is the classical density, $\Psi_\varepsilon(\mathbf{r})$ denotes the position eigenfunctions of electrons. Due to their wave nature, electrons tend to stay away from the boundaries and accumulate into more free regions inside their confinement domain. As a result, the local electron density is almost zero near to the boundaries of the shell structure, whereas it is higher in the regions distant from the boundaries. Blue strips that are formed near to boundaries are called the QBLs and they have a thickness ($\delta$) in the order of thermal wavelength of electrons, $\delta\propto \lambda_{th}$ \cite{uqbl}. When Fermi statistics is considered, QBL thickness has also inverse square root dependence on the degeneracy, $\delta\propto \lambda_{th}/\sqrt{\Lambda(T)}$. The region outside of QBLs is called the effective volume. QBLs of inner (core) and outer (material) boundaries can overlap which changes the effective volume. The amount of QBL overlap depends on the configuration angle $\theta$, which makes the effective volume shape-dependent. As a direct consequence of Fermi statistics, electron density exhibits Friedel oscillations which are at their strongest in the most degenerate case. 

Due to the strong confinement effect, in all cases, few transverse energy eigenstates determine the shape-dependent thermodynamic properties of the system. In Fig. 3(b), we show the Fermi distribution (red curve) by denoting the first 50 energy eigenstates (black dots) for the most degenerate condition. Fermi energy (in the units of $k_BT$) is as high as 74. There are several states lying directly on the Fermi surface and variations in the distributions of those states are the direct causes of entropy and heat capacity oscillations. In addition to this, effective volume and Fermi level also fluctuate up and down with changing $\theta$, which generates additional quantum oscillations to the all thermodynamic properties as they all depend on the chemical potential $\mu$.

Another important feature of the transverse energy states is their level degeneracy. Ground state and excited low-lying states are four-fold degenerate (except for small $\theta$), as dictated by the characteristic geometry of the confinement domain. This can clearly be seen in Fig. 3(c), where we plot the number of states (NOS) with respect to energy (in reverse axes). Red line denotes the Fermi level. There are 28 states below the Fermi level and all of them have four-fold degeneracy. Therefore, we would expect 7 number of Friedel oscillation peaks in each quarter slice of the confinement domain and this is exactly what is observed in Fig. 3(a). Each triangular region has 7 density peaks, which makes in total 28 peaks for 28 states. This observation is consistent in other cases and angular configurations as well.

In Fig. 3(d), quantum thermal density of electrons is plotted for the least degenerate case. At first, the result looks surprising because the density distribution looks pretty much like the ground state distribution. This is because only ground states of transverse modes substantially contribute to the physical properties for the chosen temperature and density parameters. As is seen in Fig. 3(f), only four-fold degenerate ground states lie below the Fermi level. The system is so weakly degenerate that it can be described by just a few states in the transverse direction. The distribution of the first 10 energy eigenstates is given in Fig. 3(e). Because of very weak degeneracy, the thermal distribution of states no longer looks like Fermi statistics but resembles more to the Boltzmann statistics. Indeed, under these conditions, statistical properties of the system almost obey the Boltzmann statistics. As expected, we do not observe any sign of density oscillations in the least degenerate case. Here we should also note that, despite the shape dependence being determined by a few transverse eigenstates, the longitudinal modes are abundant so that the system has a large number of particles in all cases.

The reason for the stronger magnitude of QShE in weakly degenerate systems can also directly be seen by comparing Fig. 3(a) and (d). In the least degenerate case, QBLs are much thicker than the most degenerate case. The thicker the QBLs, the larger their overlaps and stronger the QShE. The variations in thermodynamic quantities are smooth in weakly degenerate cases, because overlaps of QBLs do not exhibit oscillations with shape. In the degenerate cases, however, because of the oscillatory variations in the overlaps of QBLs (due to Friedel oscillations), effective volume (and effective density) also oscillates, which is the direct reason of the oscillatory changes in chemical potential and Fermi energy. 

In Ref. \cite{Dai_2004}, the authors discuss the importance of interactions in comparison to the boundary effects. Their analysis constitutes a base also for QShE. They compare the corrections for the chemical potential of a confined Fermi gas due to both boundary effects and interactions. It has been shown that the interactions remain negligible as long as the average system size (double volume over the surface area) $L_g=2V/S$ and particle density $n$ are small enough such that $L_g n^{2/3}<<1/a$ (the condition in Eq. (65) of Ref. \cite{Dai_2004} is reexpressed) where $a$ is the scattering length for the interactions. An approximate and simple calculation shows that $a$ is in the order of 0.6-1 nm for electron-electron (e-e) interactions in GaAs for the density and temperature values we consider. Therefore, the worst-case occurs for the highest electron density $n=2\times 10^{24}$ m$^{-3}$ which gives the condition of $L_g<<105$ nm. If we calculate (in our case $L_g=2A/P$ for a 2D core-shell plane) $L_g=(L_s-L_c)/2$, we obtain $L_g=11.5$ nm $<<105$ nm. In other words, interactions do not cause a considerable change in the results. This is an expected result due to low electron densities. A similar calculation shows that the effect of electron-phonon (e-ph) interactions is also negligible both because of low electron density and low temperatures causing low phonon densities. Furthermore, boundary modifications, induced by the change in orientation angle of the core structure, directly affect the energy spectrum of confined electrons and thus the total kinetic energy determining the Fermi level while these changes in energy spectrum indirectly and slightly effects the total interaction energy due to e-e and e-ph interactions which are already small in comparison to total kinetic energy.

\subsection{Masking of the temperature effect}
In non-degenerate cases, stronger quantum shape dependence occurs at lower temperatures due to the longer de Broglie wavelength of particles. However, this is not always the case in degenerate conditions. As seen in Table I, the Fermi wavelengths are even slightly longer at high temperatures for a constant density. This is because the increment of the temperature in a degenerate Fermi gas causes the decrease of degeneracy and, hence, indirectly contributes to lower the average energy of the particles. At the same time, however, the rise of temperature directly contributes to increasing the particles' energy. Therefore, the magnitude of QShE non-trivially depends on the changes in temperature. For the ranges in Table I, the indirect mechanism is dominant and the particles have longer Fermi wavelengths for higher temperatures. This explains why we obtain larger QShE at higher temperatures corresponding to the same electron density for some quantities; compare the blue and red solid curves in Fig. 2(a). On the other hand, the oscillation magnitudes are sensitive to the changes in the electron density, so that the density has a more dominant control on oscillations in comparison with that of the temperature. To examine this behavior in detail, we compare the relative change of QShE in the specific heat with respect to the electron density and temperature in Fig. 4. The larger the deviations from the gray plane (denoting the zero QShE), the larger the QShE. The trend behavior of larger QShE at lower temperatures and densities is clearly seen in Fig. 4. However, at the same time, this behavior does not strictly take place at all places of the density-temperature space, because the magnitude of the oscillations may locally violate this trend. For example, around $n=15\times 10^{23}$ m$^{-3}$, the magnitude of the QShE rises locally with increasing temperature. Essentially, the magnitude of QShE is not always at its maximum in case of the ratio for $\theta=0^{\circ}$ and $\theta=45^{\circ}$, due to the oscillations. Nonetheless, Fig. 4 provides a convenient way to understand the reason for the occurrence of larger QShE at higher temperatures in some regions of density-temperature space.

\subsection{Breaking of entropy-heat capacity equivalence}
Entropy and heat capacity of non-interacting, unconfined Fermi gases are equal to each other in the completely degenerate limit ($\Lambda\xrightarrow{}\infty$) \cite{pathbook} and almost equal in degenerate case. This equivalence can be broken due to quantum size effects \cite{aydin5}. Here, we investigate the fate of this equivalence under QShE and find that the entropy-heat capacity ratio of degenerate electrons deviates from unity in an oscillatory fashion. In Fig. 5, $S/C_{W}$ ratio is plotted for the $\theta$ ranging between $0^{\circ}$ and $45^{\circ}$, and density ranging from $10^{23}$ m$^{-3}$ to $2\times 10^{24}$ m$^{-3}$. The gray plane denotes unity for the sake of easy comparison. The oscillations are denser at $20$K (Fig. 5a) compared to $50$K (Fig. 5b) due to the stronger degeneracy in the former. The violation of the entropy-heat capacity equivalence is quite strong (changes up to $40\%$) in the considered ranges of shape-density-temperature values. In fact, the violation becomes more substantial with increasing degeneracy. Quantum shape dependence distinctively affects entropy and heat capacity. As a result, the breaking of their equivalence exhibits irregular oscillations.

\subsection{Controlling Fermi energy by shape}
Tuning the energy levels and controlling the Fermi level via doping, gating and size effects provide rich ways of diversification of the material properties \cite{Masumoto_1997,Macks_2000,Li_2001,Li_2006,Avetisyan_2009,Bryan_2014,Zhang_2017,Warren_2019}. Here we provide the existence of the possibility of tuning the Fermi level by shape alone, without the size effects to be engaged in. Quantum shape effect exists not only at finite temperatures, but also at the zero temperature. In Fig. 6, we examine the shape dependence of the Fermi energy of electrons at various densities. We show the relative difference of the Fermi energy from its reference value of zero degree angular configuration. It is clearly seen that changing the configuration angle of the core structure changes the Fermi energy, making it an explicit function of $\theta$. The oscillatory behavior appears in the relative variation of the Fermi energy for higher electron densities while the changes become smooth variations for lower densities with a larger magnitude. The shift in Fermi energy due to shape change is always negative for weakly degenerate densities while it oscillates around zero and takes both positive and negative values for a higher density ($2\times 10^{24}$ m$^{-3}$). For intermediate density ($10^{24}$ m$^{-3}$), however, it becomes strictly positive. These behaviors are directly related with the characteristic oscillations of the chemical potential, as Fig. 6 can be compared to Fig. 2(a) and (b) for obvious similarities. Variation of the effective volume with shape changes the system's Fermi level accordingly because of the fixed constraints of electron density and temperature. The shape control on the Fermi energy may provide a novel mechanism for a fine tuning of the position of Fermi level in band structure and changing the polarity of semiconductors. In other words, QShE causes shape-induced doping and we can finely adjust the Fermi level just by changing the configuration angle of the core structure without additional material doping. In this way, for example, we can change the polarity of the semiconductor. This may allow us to design and produce shape-induced and finely tuned junctions for solar cells, LEDs and other nanoelectronic devices. By using a twisted core structure, we can change the Fermi level and the polarity of the shell semiconductor continuously through the longitudinal direction of the structure and create a junction just by changing the orientation angle of the core.

\begin{figure}[t]
\centering
\includegraphics[width=0.48\textwidth]{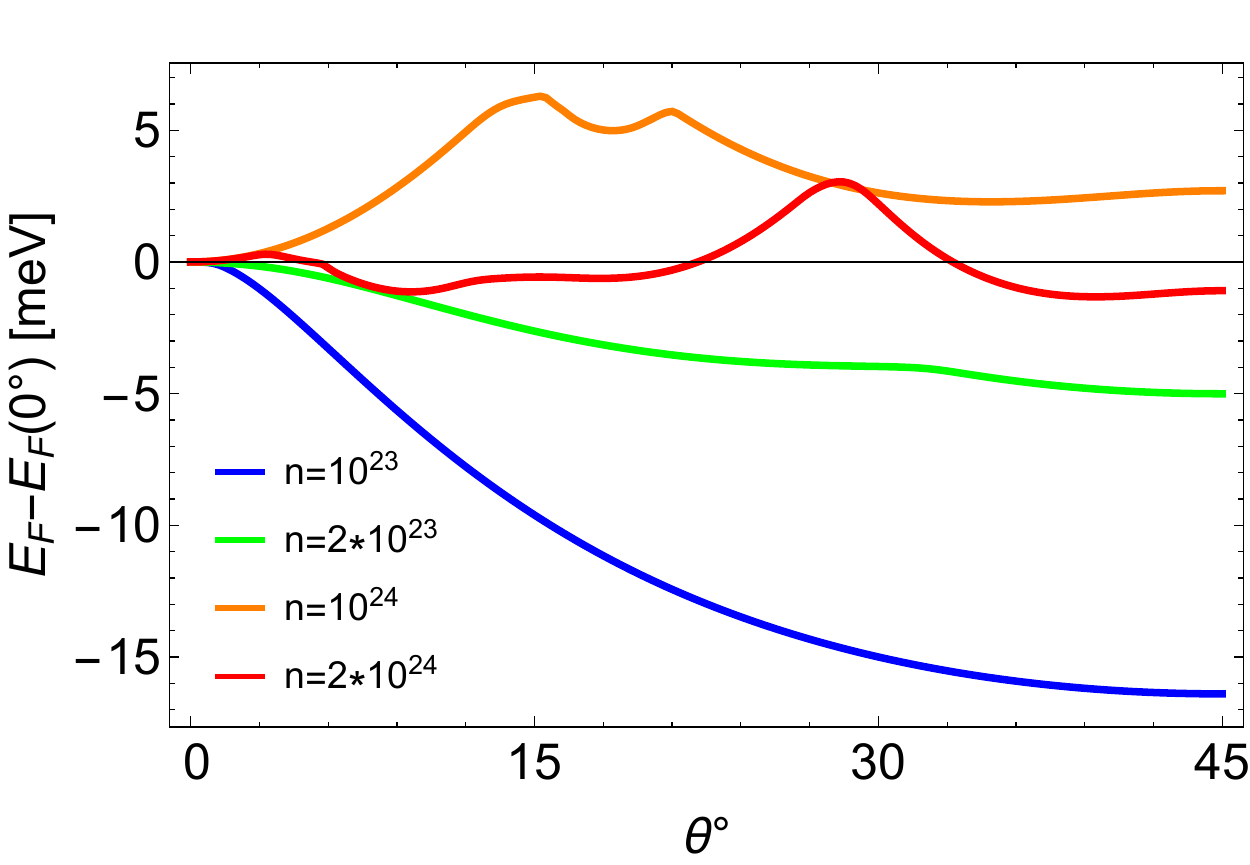}
\caption{Shape dependence of Fermi energy for different electron densities. Fermi energy is plotted relative to the reference $E_F$ at $\theta=0^{\circ}$ angular configuration.}
\label{fig:pic6}
\end{figure}

\section{Conclusion}\label{IV}
In this work, we have shown the existence of shape dependent quantum oscillations in thermodynamic properties of strongly confined degenerate electrons due to QShE. QShE can arise when electrons are confined in nested nanostructures like the core-shell nanowires considered here. Unlike the quantum size oscillations, we observe oscillations in the chemical potential and internal energy due to changes in shape for moderately degenerate electron densities. As expected, the entropy and heat capacity also show oscillating behaviors. However, the origins of these two types of oscillations are not the same. While the chemical potential and internal energy oscillations depend on the fluctuations of the Fermi level and hence the occupancy function itself, the entropy and heat capacity oscillations result from the fluctuations of both the occupancy variance and the Fermi level. The former shape oscillations directly depend on the changes in effective volume. The latter ones depend both on effective volume and fluctuations of states around the Fermi surface. QBL concept and transverse energy eigenvalue analyses provide even deeper physical insights on the existence and explanation of quantum shape oscillations. The frequency of Friedel oscillations can also be determined from the number of degenerate ground and low-lying transverse eigenstates.

We notice that QShE causes appreciable changes in thermodynamic properties of confined electrons and changes can rise up to $60\%$ for the considered ranges of parameters. Our results suggest QShE are not negligible and should be taken into account when appropriate, e.g. under strong confinements leading QBLs to overlap. It is also seen that the Fermi energy can be finely tuned by shape effects providing a mechanism to control the Fermi energy and the polarity of semiconductors. This work constitutes to be the first study on shape dependent oscillations in the thermodynamic properties of particles obeying Fermi-Dirac statistics.

Further studies based on more complicated models for QShE on the band structure of semiconductors are needed as an extension of the proposed idea and its results. Differences in longitudinal electronic properties of 2D structures due to the difference in configuration angle of the core region may be a useful extension of the study. Another interesting extension can be the consideration of twisted core-shell semiconductor structures and examining the changes in their electronic properties through the longitudinal direction. Such a kind of twisted semiconductor core-shell structure is expected to exhibit Peltier effect due to the shape-induced changes in polarity along the longitudinal direction, if we try to keep it at a constant temperature while an electric current passes through.

The results represent the maximum possible quantum shape oscillations on thermodynamic properties of confined electrons in ideal core-shell GaAs structures. Therefore, consideration of the effects of defects in shell structure and boundary roughness as well as the interactions on quantum shape oscillations can be other extensions of the study.

\acknowledgments{The authors would like to thank Oscar Gr\aa n\"as for the useful discussions on the comparison of DFT and non-interacting electron models. JF acknowledges the support from Vetenskapsr\aa det. AS expresses his deep gratitude to the Material Theory Division, Department of Physics $\&$ Astronomy, Uppsala University for their kind support and warm hospitality.}

\bibliography{shfdref}
\bibliographystyle{unsrt}
\end{document}